\documentclass[aps,prl,superscriptaddress,twocolumn,showpacs]{revtex4}
\usepackage{graphics}
\usepackage{amsmath}
\usepackage{graphicx}
\usepackage{amsfonts}
\usepackage{amssymb}

\begin{document}

\title{Direct and reverse secret-key capacities of a quantum channel}
\author{Stefano Pirandola}
\affiliation{M.I.T. - Research Laboratory of Electronics, Cambridge MA 02139, USA}
\author{Raul Garc\'{\i}a-Patr\'{o}n}
\affiliation{M.I.T. - Research Laboratory of Electronics, Cambridge MA 02139, USA}
\author{Samuel L. Braunstein}
\date{\today }
\affiliation{Computer Science, University of York, York YO10 5DD, United Kingdom}
\author{Seth Lloyd}
\affiliation{M.I.T. - Research Laboratory of Electronics, Cambridge MA 02139, USA}
\affiliation{M.I.T. - Department of Mechanical Engineering, Cambridge MA 02139, USA}

\begin{abstract}
We define the direct and reverse secret-key capacities of a memoryless
quantum channel as the optimal rates that entanglement-based quantum key
distribution protocols can reach by using a \textit{single} forward
classical communication (direct reconciliation) or a \textit{single}
feedback classical communication (reverse reconciliation). In particular,
the reverse secret-key capacity can be positive for antidegradable channels,
where no forward strategy is known to be secure. This property is explicitly
shown in the continuous variable framework by considering arbitrary one-mode
Gaussian channels.
\end{abstract}

\maketitle

Since the birth of quantum information \cite{Nielsen}, both the notions of
quantum entanglement and memoryless quantum channel have been fundamental in
many theoretical investigations. This consideration is particularly true in
the field of quantum cryptography. On the one hand, entanglement
distribution is a basic process in the formulation of quantum key
distribution (QKD) protocols \cite{Intro}. On the other hand, memoryless
quantum channels can be seen as the effect of collective attacks, recognized
as predominant in quantum cryptography after the recent achievements of Ref.~%
\cite{Renner}. In this paper, we consider a generic QKD protocol where two
honest parties (Alice and Bob) extract a secret key from the remote
correlations that are generated by one of the parties (Alice) after the
distribution of a generic entangled state over a memoryless quantum channel.
Such a task can be assisted by one-way classical communications (CCs) which
can be forward, i.e., from Alice to Bob, or feedback, i.e., from Bob to
Alice. Even if the scenario can seem symmetric, it is actually much harder
to study the feedback-assisted protocols and optimize the corresponding
secret-key rates. The reason relies on the fact that Alice can actively
exploit the information already received from Bob for conditioning the
subsequent inputs to the quantum channel. In this paper we simplify this
problem by restricting the feedback to a \textit{single} (and therefore
final) CC from Bob. Although we make this restriction on the feedback
strategy, the security performance is still remarkable. Under suitable
conditions, the corresponding QKD\ protocols are in fact able to outperform
all the known QKD protocols which are based on forward CCs.

In general, we identify the notions of direct and reverse reconciliation
\cite{RR} with the ones of assistance by a \textit{single} forward and a
\textit{single} feedback CC, respectively. Then, by optimizing over
corresponding protocols, we define the direct and reverse secret-key
capacities of a quantum channel. In direct reconciliation, the optimization
over a single forward CC is not restrictive at all. In fact, the direct
secret-key capacity represents an equivalent entanglement-based formulation
of the (forward) secret-key capacity of Ref.~\cite{Privacy}. In reverse
reconciliation, even if the feedback strategy is limited, the security
performance is in any case outstanding. In fact, the reverse secret-key
capacity can be positive even if the quantum channel is antidegradable \cite%
{DevShor}, i.e., an eavesdropper is able to reconstruct completely the
output state of the receiver (and no forward protocol is known to be
secure). This property is explicitly shown in the most important scenario
for the continuous variable QKD: the one-mode Gaussian channel. In order to
establish this result in its full generality, we resort to the recent
canonical classification of the one-mode Gaussian channels \cite%
{HolevoCanonical} (see also Refs.~\cite{HETpaper,AlexJens}). By assuming an
arbitrary one-mode Gaussian channel, we then exploit a general lower bound
for the reverse secret-key capacity. This bound corresponds to the additive
capacity of Ref.~\cite{Raul}, which is connected to the entanglement
distillation by feedback CCs. For one-mode Gaussian channels, this additive
capacity assumes an analytical formula which is exceptionally simple. Most
importantly, it enables us to prove the positivity of the reverse secret-key
capacity over a wide range of parameters where the channel is
antidegradable. In a final investigation, we also prove that tighter bounds
can be derived by exploiting noise effects in the key-distribution process.

\begin{figure}[tbph]
\vspace{-2.0cm}
\par
\begin{center}
\includegraphics[width=0.5\textwidth] {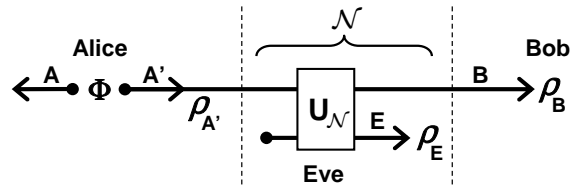}
\end{center}
\par
\vspace{-2.3cm}
\caption{Quantum channel $\mathcal{N}$\ and its dilation.}
\label{CHANNELpic}
\end{figure}

Let us consider an arbitrary quantum channel, i.e., a completely positive
trace-preserving (CPT) map $\mathcal{N}$, transforming the input state $\rho
_{A^{\prime }}$ of a sender (Alice) into the output state $\rho _{B}$\ of a
receiver (Bob). As depicted in Fig.~\ref{CHANNELpic}, such a channel can
always be represented by\ an isometric embedding $U_{\mathcal{N}}:\mathcal{H}%
_{A^{\prime }}\rightarrow \mathcal{H}_{B}\otimes \mathcal{H}_{E}$ followed
by a trace over the environment $E$\ which we identify with the eavesdropper
(Eve). By definition, the original channel $\mathcal{N}$ is called \textit{%
degradable} if there exists a CPT map $\mathcal{D}$ such that $\rho _{E}=%
\mathcal{D}(\rho _{B})$, where $\rho _{E}$ is the output state of Eve. By
contrast, $\mathcal{N}$ is called \textit{antidegradable} if there exists a
CPT map $\mathcal{\tilde{D}}$ such that $\rho _{B}=\mathcal{\tilde{D}}(\rho
_{E})$. A further dilation of the quantum channel is provided by the
purification of the input $\rho _{A^{\prime }}=\mathrm{Tr}_{A}\Phi $ with $%
\Phi :=\left\vert \Phi \right\rangle \left\langle \Phi \right\vert
_{AA^{\prime }}$, involving the introduction of a supplementary system $A$
at Alice's side. Notice that the three output systems $A,B$ and $E$ are
globally described by a pure state $\Psi =\left\vert \Psi \right\rangle
\left\langle \Psi \right\vert _{ABE}$, which is given by $\Psi
=(I_{A}\otimes U_{\mathcal{N}})(\Phi )$. A key-distribution protocol can be
introduced by extending the scenario of Fig.~\ref{CHANNELpic} to $n$\
(entangled) uses of the channel and by adding measurements to Alice's and
Bob's sides. Restricting the honest users to a single one-way CC, we have a
one-way key-distribution protocol which can be direct, if Alice assists Bob,
or reverse, if Bob assists Alice.

\begin{figure}[tbph]
\vspace{-2.4cm}
\par
\begin{center}
\includegraphics[width=0.55\textwidth] {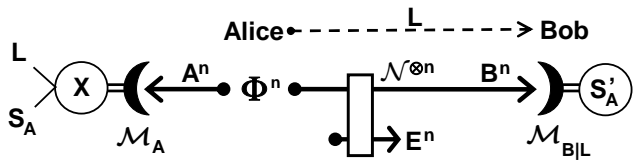}
\end{center}
\par
\vspace{-2.7cm}
\caption{Key-distribution protocol in direct reconciliation.}
\label{DRpic}
\end{figure}

Let us begin with the \textit{direct protocol}. In the first step of this
protocol, Alice distributes a pure entangled state $\Phi ^{n}:=\left\vert
\Phi \right\rangle \left\langle \Phi \right\vert _{A^{n}A^{\prime n}}$
sending the $A^{\prime }$-part through the memoryless quantum channel $%
\mathcal{N}^{\otimes n}=\mathcal{N}\otimes \cdots \otimes \mathcal{N}$ (see
Fig.~\ref{DRpic}). As a consequence, we have a pure state $\Psi
^{n}=(I_{A}\otimes \hat{U}_{_{\mathcal{N}}})^{\otimes n}(\Phi ^{n})$, shared
by the output systems of Alice ($A^{n}$), Bob ($B^{n}$) and Eve ($E^{n}$).
On her local systems $A^{n}$, Alice performs a quantum measurement $\mathcal{%
M}_{A}$. This is generally described by a positive operator-valued measure
(POVM) $\{\hat{A}_{x}\}$ with outcomes $x$. As a result, she gets an output
random variable $X=\{x,p(x)\}$ where the values $x$ have probability
distribution $p(x)=\mathrm{Tr}(\Psi ^{n}\hat{A}_{x})$. After the
measurement, Alice processes $X$ via a classical channel $X\rightarrow
(S_{A},L)$, which yields a key variable $S_{A}=\{s,p(s)\}$ and an assisting
variable $L=\{l,p(l)\}$. The assisting variable $L$ contains all the
information necessary to Bob for performing error correction and privacy
amplification. The value $l$\ of $L$ is then broadcast by Alice through a
public channel. Using this information, Bob performs a conditional POVM $%
\mathcal{M}_{B|L}=\{\hat{B}_{s}^{(l)}\}$ on his systems $B^{n}$, and
retrieves an estimation of Alice's key $S_{A}^{\prime }$ up to an error
probability $p(S_{A}\neq S_{A}^{\prime })\leq \varepsilon $ \cite{NotePRACT}.

In the limit of $n\rightarrow +\infty $, Alice and Bob are able to rule out
Eve completely and share exactly the same uniform key $S_{A}$ corresponding
to $H(S_{A}):=nR$ secret bits. The highest secret-key rate $R$ which is
achievable by direct protocols over a quantum channel $\mathcal{N}$ is
called the \textit{direct secret-key capacity }$K_{\blacktriangleright }(%
\mathcal{N})$ of the channel. This quantity is characterized by the formula
\cite{sup,Details}%
\begin{equation*}
K_{\blacktriangleright }(\mathcal{N})=\lim_{n\rightarrow \infty }\frac{1}{n}%
\max_{\substack{ \Phi ^{n},\mathcal{M}_{A}  \\ X\rightarrow T}}\left[
I(X:B^{n}|T)-I(X:E^{n}|T)\right] ~,
\end{equation*}%
where the maximum is over all the pure states $\Phi ^{n}$, Alice's POVMs $%
\mathcal{M}_{A}$, and all the classical channels $p(t|x):X\rightarrow T$
generating the conditioning dummy variable $T$. In this formula, $%
I(X:B^{n}|T)$ and $I(X:E^{n}|T)$ are the conditional Holevo information of
Bob and Eve \cite{state2}. In direct reconciliation, one can show \cite%
{Details} that the conditioning by $T$ can actually be avoided in the
previous formula, and that $K_{\blacktriangleright }(\mathcal{N})$ is the
entanglement-based version of the secret-key capacity $K(\mathcal{N})$ of
Ref.~\cite{Privacy}. From Ref.~\cite{Privacy} it is known that $K(\mathcal{N}%
)\geq E(\mathcal{N})=Q(\mathcal{N})$, where $E(\mathcal{N})$ is the
entanglement-generation capacity of $\mathcal{N}$ and $Q(\mathcal{N})$ its
unassisted quantum capacity \cite{SethLSD}.

\begin{figure}[tbph]
\vspace{-2.4cm}
\par
\begin{center}
\includegraphics[width=0.55\textwidth] {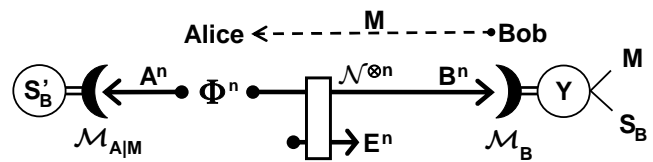}
\end{center}
\par
\vspace{-2.7cm}
\caption{Key-distribution protocol in reverse reconciliation.}
\label{RRpic}
\end{figure}

A key-distribution protocol in reverse reconciliation, i.e., a \textit{%
reverse protocol}, consists of interchanging Alice and Bob in terms of the
assisting CC while keeping Alice as dispenser of the quantum state $\Phi
^{n} $. As we have already mentioned, this is a particular case of a more
general feedback-assisted protocol, where Alice distributes the entangled
state $\Phi ^{n}$ in $n$ different rounds, each one conditioned by previous
CCs received from Bob. In the first step of a reverse protocol, Alice
distributes $\Phi ^{n}$ generating $\Psi ^{n}$ as before (see Fig.~\ref%
{RRpic}). But now the first measurement is done by Bob, who detects his
systems $B^{n}$ via a POVM $\mathcal{M}_{B}$, generating an output variable $%
Y=\{y,p(y)\}$. Again, this variable is processed into a key variable $S_{B}$
and an assisting variable $M$, which is broadcast by Bob. Using this
information, Alice subjects her local systems $A^{n}$ to a conditional POVM $%
\mathcal{M}_{A|M}$, retrieving an estimation of Bob's key $S_{B}^{\prime }$
up to a small error probability. The \textit{reverse secret-key capacity} $%
K_{\blacktriangleleft }(\mathcal{N})$ of a quantum channel $\mathcal{N}$ is
defined as the highest secret-key rate which is achievable by reverse
protocols over $\mathcal{N}$. For this capacity we can prove the upper bound
\cite{sup,Details}%
\begin{equation*}
K_{\blacktriangleleft }(\mathcal{N})\leq \lim_{n\rightarrow \infty }\frac{1}{%
n}\max_{\substack{ \Phi ^{n},\mathcal{M}_{B}  \\ Y\rightarrow T}}\left[
I(Y:A^{n}|T)-I(Y:E^{n}|T)\right] ~,
\end{equation*}%
where the maximum is now over Bob's POVMs $\mathcal{M}_{B}$ and involves the
processing of Bob's variable $Y$.

In order to find achievable lower bounds for this capacity, let us restrict
the process of key-distribution to the one of key-distillation. A one-way
\textit{key-distillation} protocol over a channel $\mathcal{N}$ (in direct
or reverse reconciliation) is defined as a one-way key-distribution protocol
where the input state is separable over different uses of the channel, i.e.,
$\Phi ^{n}=\Phi ^{\otimes n}$. Maximizing over these protocols, we can
define the direct and reverse \textit{key-distillation capacities} of a
quantum channel, that we denote by $K_{\blacktriangleright }^{\otimes }(%
\mathcal{N})$ and $K_{\blacktriangleleft }^{\otimes }(\mathcal{N})$. These
capacities clearly satisfy $K_{\blacktriangleright }^{\otimes }(\mathcal{N}%
)\leq K_{\blacktriangleright }(\mathcal{N})$ and $K_{\blacktriangleleft
}^{\otimes }(\mathcal{N})\leq K_{\blacktriangleleft }(\mathcal{N})$. Using
the results of Ref.~\cite{DW}, we can easily prove the formulas~\cite%
{sup,Details}%
\begin{eqnarray*}
K_{\blacktriangleright }^{\otimes }(\mathcal{N}) &=&\lim_{n\rightarrow
\infty }\frac{1}{n}\max_{_{\substack{ \Phi ,\mathcal{M}_{A}  \\ X\rightarrow
T }}}\left[ I(X:B^{n}|T)-I(X:E^{n}|T)\right] ~, \\
K_{\blacktriangleleft }^{\otimes }(\mathcal{N}) &=&\lim_{n\rightarrow \infty
}\frac{1}{n}\max_{\substack{ \Phi ,\mathcal{M}_{B}  \\ Y\rightarrow T}}\left[
I(Y:A^{n}|T)-I(Y:E^{n}|T)\right] ~,
\end{eqnarray*}%
where now the maximization is over the single copy\ of the state $\Phi $,
i.e., over $\Phi ^{\otimes n}$. Exploiting the relation with the
key-distillation, we can prove an important lower bound for $%
K_{\blacktriangleleft }(\mathcal{N})$. In fact, we have \cite{Details}%
\begin{equation}
K_{\blacktriangleleft }(\mathcal{N})\geq K_{\blacktriangleleft }^{\otimes }(%
\mathcal{N})\geq E_{R}^{(1)}(\mathcal{N})=E_{R}(\mathcal{N})~,
\label{bounds}
\end{equation}%
where $E_{R}(\mathcal{N})$ is the additive capacity of Ref.~\cite{Raul}. In
particular, the single-letter version of this capacity is given by the
formula $E_{R}^{(1)}(\mathcal{N}):=\max_{\left\vert \Phi \right\rangle
}I(A\langle B)$, where $I(A\langle B):=H(\rho _{A})-H(\rho _{AB})$ is the
\textit{reverse coherent information} computed over Alice and Bob's output
state $\rho _{AB}=(I_{A}\otimes \mathcal{N})(\Phi )$. Remarkably, $E_{R}(%
\mathcal{N})$ has a very different behavior with respect to the quantum
capacity $Q(\mathcal{N})$. In particular, for an antidegradable channel, we
can have $E_{R}(\mathcal{N})>0$ which implies $K_{\blacktriangleleft }(%
\mathcal{N})>0$. In the following this is explicitly shown for a generic
Gaussian channel affecting a single bosonic mode.

\begin{figure}[tbph]
\vspace{-2.3cm}
\par
\begin{center}
\includegraphics[width=0.54\textwidth] {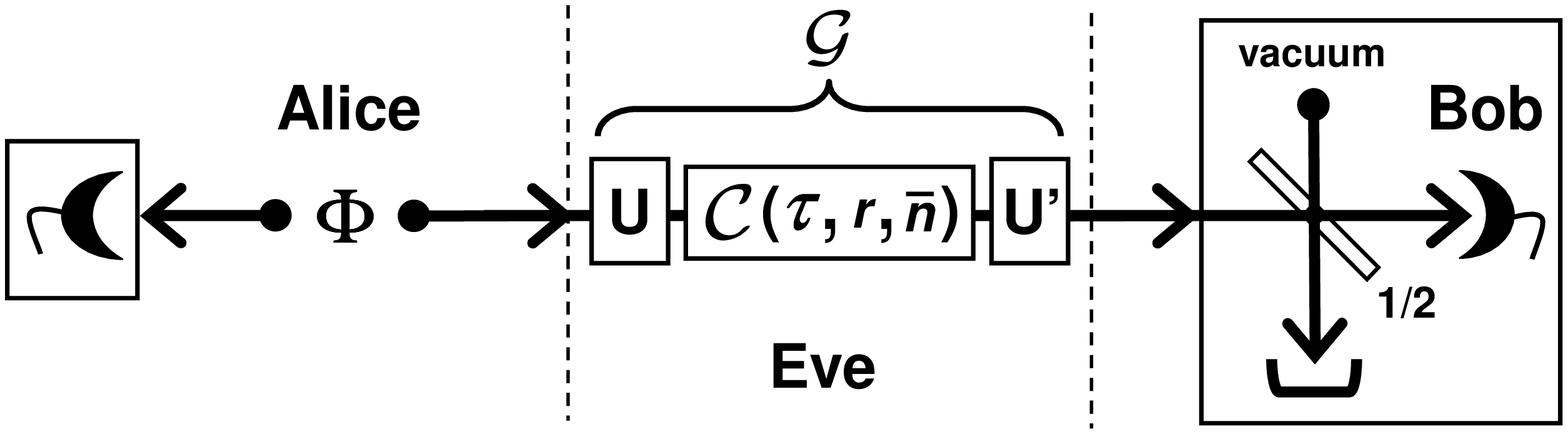}
\end{center}
\par
\vspace{-2.3cm}
\caption{\textbf{Canonical decomposition.} In the center of the figure
(Eve), we show the decomposition of a one-mode Gaussian channel $\mathcal{G}$%
\ into a canonical form $\mathcal{C}(\protect\tau ,r,\bar{n})$\ up to a pair
of unitaries $\hat{U}$ and $\hat{U}^{\prime }$. \textbf{Noisy reverse
protocol.}~The reverse protocol with the rate of Eq.~(\protect\ref{rateOUT})
is achieved by applying two inverse unitaries $\hat{U}^{-1}$ and $\hat{U}%
^{\prime -1}$ (not shown in the figure), restricting the quantum
distribution to a two-mode squeezed vacuum state $\Phi =$ $\left\vert
\protect\mu \right\rangle \left\langle \protect\mu \right\vert $, and
providing Alice and Bob with homodyne detectors (see boxes in the figure).
In particular, Bob's homodyne detector is placed after one of the two output
ports of a balanced beam-splitter mixing the signal with the vacuum (the
output of the other port is discarded).}
\label{ProtPIC}
\end{figure}

\begin{figure}[tbph]
\vspace{+0.2cm}
\par
\begin{center}
\includegraphics[width=0.42\textwidth] {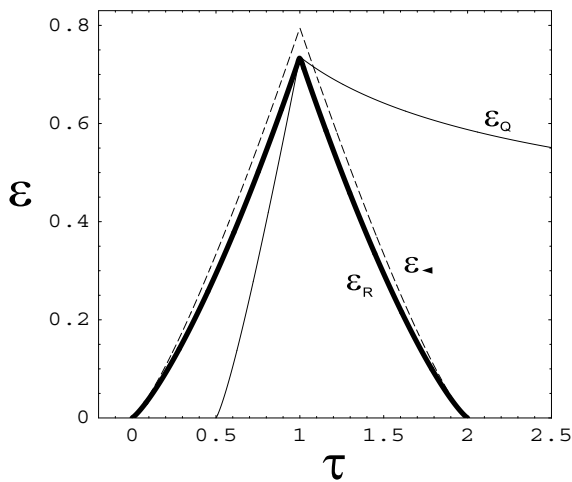}
\end{center}
\par
\vspace{-0.5cm}
\caption{Scaled thermal noise $\protect\varepsilon $ versus transmission $%
\protect\tau \neq 1$. The thin solid curve $\protect\varepsilon _{Q}=\protect%
\varepsilon _{Q}(\protect\tau )$ refers to $Q^{(1,g)}(\mathcal{G})$, while
the thick solid curve $\protect\varepsilon _{R}=\protect\varepsilon _{R}(%
\protect\tau )$ refers to $E_{R}(\mathcal{G})$. Above (below) these curves
the corresponding capacities are zero (positive). Notice that $E_{R}(%
\mathcal{G})$ is positive in the region $0<\protect\tau \leq 1/2$ and $%
\protect\varepsilon <\protect\varepsilon _{R}(\protect\tau )$, where the
channel is antidegradable. The dashed curve $\protect\varepsilon %
_{\blacktriangleleft }=\protect\varepsilon _{\blacktriangleleft }(\protect%
\tau )$ corresponds to $R_{\blacktriangleleft }=0$, where $%
R_{\blacktriangleleft }$ is the rate given in Eq.~(\protect\ref{rateOUT}).}
\label{PlotSecurity}
\end{figure}

Recall that a bosonic mode is a quantum system described by a pair of
quadrature operators, $\hat{q}$ and $\hat{p}$, with $[\hat{q},\hat{p}]=2i$.
Then, a Gaussian channel $\mathcal{G}$ acting on this system is a CPT map
which preserves the Gaussian statistics of its states. Using the compact
formulation of Ref.~\cite{HETpaper}, every $\mathcal{G}$ can be associated
with three symplectic invariants: \textit{transmission} $\tau $, \textit{rank%
} $r$, and \textit{temperature} $\bar{n}$. These invariants completely
characterize the unique canonical form \cite{HolevoCanonical} $\mathcal{C}%
(\tau ,r,\bar{n})$ which is unitarily equivalent to $\mathcal{G}$ (see Fig.~%
\ref{ProtPIC}). For a generic one-mode Gaussian channel $\mathcal{G}$ with
transmission $\tau \neq 1$, we compute \cite{Details}%
\begin{equation}
E_{R}(\mathcal{G})=\max \left\{ 0,\log \left\vert \frac{1}{1-\tau }%
\right\vert -g(\bar{n})\right\} ~,  \label{Q_revG}
\end{equation}%
where $g(x):=(x+1)\log (x+1)-x\log x$. This expression must be compared with
\begin{equation}
Q^{(1,g)}(\mathcal{G})=\max \left\{ 0,\log \left\vert \frac{\tau }{1-\tau }%
\right\vert -g(\bar{n})\right\} ~,  \label{Q_directG}
\end{equation}%
which is the quantum capacity $Q(\mathcal{G})$ restricted to a single use of
the channel and pure Gaussian states \cite{HolWerner}. It is known that $Q(%
\mathcal{G})=Q^{(1,g)}(\mathcal{G})$ for a degradable channel \cite{Wolf},
while $Q(\mathcal{G})=0$ for an antidegradable channel. In order to analyze
and compare the previous quantities, we introduce the scaled thermal noise $%
\varepsilon :=2\bar{n}\left\vert 1-\tau \right\vert $. For every $\tau \neq 1
$, we then consider the minimal noises, $\varepsilon _{Q}$ and $\varepsilon
_{R}$, above which we have $Q^{(1,g)}(\mathcal{G})=0$ and $E_{R}(\mathcal{G}%
)=0$, respectively. The corresponding threshold curves $\varepsilon
_{Q}=\varepsilon _{Q}(\tau )$ and $\varepsilon _{R}=\varepsilon _{R}(\tau )$%
\ are shown in Fig.~\ref{PlotSecurity}. For $\tau \leq 1/2$, the one-mode
Gaussian channel is known to be antidegradable \cite{HolevoCanonical} and,
therefore, we have $Q(\mathcal{G})=0$. In this case, no forward protocol is
known to be secure, i.e., it is not known if $K(\mathcal{G}%
)[=K_{\blacktriangleright }(\mathcal{G})]\neq 0$. However, for $0<\tau \leq
1/2$ and $\varepsilon <\varepsilon _{R}(\tau )$, we have a wide region of
antidegradability where $E_{R}(\mathcal{G})>0$ and, therefore, $%
K_{\blacktriangleleft }(\mathcal{G})>0$. In other words, even if Eve can
reconstruct Bob's state, Alice and Bob are still able to extract a secret
key using a reverse protocol. This result is a remarkable feature of the
reverse reconciliation, which is here stated in its full generality by
considering arbitrary one-mode Gaussian channels.

As a final investigation, we prove an effective separation between $%
K_{\blacktriangleleft }$ and $E_{R}$, i.e., the existence of tighter lower
bounds for $K_{\blacktriangleleft }$. In particular, we show a reverse
key-distillation protocol whose rate $R_{\blacktriangleleft }$ can
outperform $E_{R}(\mathcal{G})$, so that we have $K_{\blacktriangleleft
}^{\otimes }(\mathcal{G})\neq E_{R}(\mathcal{G})$ and, therefore, $%
K_{\blacktriangleleft }(\mathcal{G})\neq E_{R}(\mathcal{G})$. This protocol
exploits a noisy decoding measurement as in Ref.~\cite{Raul2} and works as
follows. For every $\mathcal{G}$, Alice and Bob can in principle apply two
input-output unitaries that put $\mathcal{G}$ in canonical form. Assuming
this reduction, Alice distributes $n$ copies of a two-mode squeezed vacuum
state $\left\vert \mu \right\rangle \left\langle \mu \right\vert $ with
variance $\mu $ \cite{Qoptics}. At the output of the channel, Bob's
measurement setup consists of a balanced beam-splitter followed by a random
homodyne detection of $\hat{q}$ or $\hat{p}$, as shown in Fig.~\ref{ProtPIC}%
. The corresponding outcomes are classically processed to provide the two
variables $S_{B}$ and $M$. Then, Bob broadcasts the value of the assisting
variable $M$, containing also the correct sequence of $\hat{q}$ and $\hat{p}$
detections. As a consequence, Alice performs the same sequence of homodyne
detections on her systems $A^{n}$ and then applies error-correction and
privacy amplification to get her estimation of the key $S_{B}^{\prime }$. In
the limits for $n\rightarrow +\infty $ and $\mu \rightarrow +\infty $ (and
for $\tau \neq 1$), the honest users achieve the secret-key rate%
\begin{equation}
R_{\blacktriangleleft }=\max \left\{ 0,\tfrac{1}{2}\log \tfrac{\lambda }{%
\left\vert 1-\tau \right\vert }+g\left( \sqrt{\tfrac{w}{4\lambda }}-\tfrac{1%
}{2}\right) -g(\bar{n})\right\} ,  \label{rateOUT}
\end{equation}%
where $\lambda :=(\left\vert 1-\tau \right\vert +w)/(1+\left\vert 1-\tau
\right\vert w)$ and $w:=2\bar{n}+1$. Let us denote by $\varepsilon
_{\blacktriangleleft }=\varepsilon _{\blacktriangleleft }(\tau )$ the
security threshold corresponding to $R_{\blacktriangleleft }=0$. As shown in
Fig.~\ref{PlotSecurity}, there is a whole region for $0<\tau <2$ and $%
\varepsilon _{R}<\varepsilon <\varepsilon _{\blacktriangleleft }$, where $%
R_{\blacktriangleleft }>E_{R}(\mathcal{G})=0$. As a consequence, we
generally have $K_{\blacktriangleleft }^{\otimes }(\mathcal{G})\neq E_{R}(%
\mathcal{G})$ and, therefore, $K_{\blacktriangleleft }(\mathcal{G})\neq
E_{R}(\mathcal{G})$. Notice that when $\mathcal{G}$ is just a canonical
form, the previous protocol can be implemented in practice, without the help
of any quantum memory. Alice and Bob can in fact perform their detections
step-by-step and then keep only the data measured in the same basis ($\hat{q}
$ or $\hat{p}$). In this case the rate $R_{\blacktriangleleft }$ of Eq.~(\ref%
{rateOUT}) refers to the sifted-key after the basis reconciliation.

In conclusion, we have introduced the notions of direct and reverse
secret-key capacities of a quantum channel, specifying these notions for
key-distillation too. In particular, the reverse capacities $%
K_{\blacktriangleleft }$ and $K_{\blacktriangleleft }^{\otimes }$ extend the
concept of reverse reconciliation to a completely general scenario, where
this procedure must be intended as a classical assistance by means of a
single feedback CC. Such reverse capacities are lower bounded by an additive
quantity $E_{R}$, which is connected with entanglement distillation and has
been explicitly computed for one-mode Gaussian channels. For these channels,
we have shown that the property of antidegradability does not necessarily
preclude the possibility to extract a secret key. This is proven in full
generality without any restriction on the Gaussian model. In other words, we
have not restricted the one-mode Gaussian channel to any specific
description like, e.g., a beam-splitter with a thermal input. In this
general scenario, we have also shown an explicit protocol which proves an
effective separation between $K_{\blacktriangleleft }$ and $E_{R}$. In
future works, our results can be exploited for exploring the ultimate
cryptographic properties of arbitrary quantum channels.

The research of S.P. was supported by a Marie Curie Action within the 6th
European Community Framework Programme. R.G.P. and S.L. were supported by
the W. M. Keck foundation center for extreme quantum information theory
(xQIT).


\begin{thebibliography}{99}
\bibitem{NotePRACT} For simplicity, we include error correction and privacy
amplification (which correspond to a classical channel) into the decoding
POVM.
\bibitem{sup} In the formulas for $K_{\blacktriangleright }(\mathcal{N})$, $%
K_{\blacktriangleleft }(\mathcal{N})$, $K_{\blacktriangleright }^{\otimes }(%
\mathcal{N})$ and $K_{\blacktriangleleft }^{\otimes }(\mathcal{N})$ we
implicitly assume $\max x=\max \{0,x\}$.
\bibitem{Details} Detailed derivations will be presented elsewhere.
\bibitem{state2} In the formulas for $K_{\blacktriangleright }(\mathcal{N})$
and $K_{\blacktriangleright }^{\otimes }(\mathcal{N})$, the information
quantities refer to the classical-quantum state $\omega
_{XTB^{n}E^{n}}=\sum_{t,x}p(t|x)p(x)\left\vert x\right\rangle \left\langle
x\right\vert _{X}\otimes \left\vert t\right\rangle \left\langle t\right\vert
_{T}\otimes \rho _{B^{n}E^{n}}(x)$, where $\rho _{B^{n}E^{n}}(x)$ is Bob and
Eve's state conditioned to the outcome $x$. A corresponding state $\omega
_{YTA^{n}E^{n}}$ must be considered for $K_{\blacktriangleleft }(\mathcal{N}%
) $ and $K_{\blacktriangleleft }^{\otimes }(\mathcal{N})$.
\end{thebibliography}
\end{document}